\documentclass[conference,compsoc]{IEEEtran}
\def\BibTeX{{\rm B\kern-.05em{\sc i\kern-.025em b}\kern-.08emT\kern-.1667em\lower.7ex\hbox{E}\kern-.125emX}}

\usepackage{xspace} 
\usepackage{xurl}
\usepackage{amsmath,amsfonts,chemarrow,balance, graphicx, subcaption, url, multirow, graphics, booktabs, colortbl, algpseudocode, xcolor}
\newcommand{\BfPara}[1]{\vspace{1mm}{\noindent\bf#1.}\xspace\xspace}
\usepackage[linesnumbered,ruled]{algorithm2e}

\usepackage{tikz}

\usepackage{hyperref}
\usepackage[normalem]{ulem}

\usepackage[english]{babel}
\addto\extrasenglish{%
}
\hypersetup{
    pdfpagemode=pagewidth,
    plainpages=false,
    colorlinks,
    urlcolor=blue!70!black,
    linkcolor=red!70!black,
    citecolor=green!70!black,
}

\let\oldautoref\autoref
\renewcommand{\autoref}[1]{\textcolor{red!70!black}
{\oldautoref{#1}}}

\usepackage{paralist}
\usepackage{verbatim}

\usepackage{booktabs}

\usepackage[framemethod=TikZ]{mdframed}
\definecolor{rubcolor}{HTML}{7E8995}
\global\mdfdefinestyle{insightstyle}{%
backgroundcolor=rubcolor!3,
outerlinewidth=1pt,innerlinewidth=0pt,
outerlinecolor=rubcolor,roundcorner=5pt
}
\newmdenv[roundcorner=10pt, frametitle=Key Lessons, linecolor=rubcolor]{insightbox}

\usepackage{pifont}
\newcommand{\circleone}{\ding{202}\xspace}
\newcommand{\circletwo}{\ding{203}\xspace}
\newcommand{\circlethree}{\ding{204}\xspace}
\newcommand{\circlefour}{\ding{205}\xspace}

\begin{document}

\date{}

\title{\Large \bf An Explorative Study of \emph{Pig Butchering} Scams}

\author{
{\rm Bhupendra Acharya}\\
CISPA Helmholtz for Information Security\\
bhupendra.acharya@cispa.de 
\and
{\rm Thorsten Holz}\\
CISPA Helmholtz for Information Security\\
holz@cispa.de

}

\maketitle

\pagestyle{plain}

\begin{abstract}
In the recent past, so-called \emph{pig-butchering} scams are on the rise. This term is based on a translation of the Chinese term \emph{Sha Zhu Pan}, where scammers refer to victims as \emph{pig} which are to be ``fattened up before slaughter'' so that scammer can siphon off as much monetary value as possible. In this type of scam, attackers perform social engineering tricks on victims over an extended period of time to build credibility or relationships, in contrast to similar scams such as romance, cryptocurrency, investment, and job fraud. After a certain period, when victims eventually transfer larger amounts of money to scammers, the fraudsters' platforms or profiles go permanently offline and the victims' money is lost.

In this work, we provide the first comprehensive study of pig-butchering scams from multiple vantage points. Our study analyzes the direct victims' narratives shared on multiple social media platforms, public abuse report databases, and case studies from news outlets. Between March 2024 to October 2024, we collected data related to pig butchering scams from (i) four social media platforms comprised of more than 430,000 social media accounts and 770,000 posts; (ii) more than 3,200 public abuse reports narratives, and (iii) about 1,000 news articles. Through automated and qualitative evaluation, we provide an evaluation of victims of pig-butchering scams, finding 146 social media scammed users, 2,570 abuse reports narratives, and 50 case studies of 834 souls from news outlets. In total, we approximated losses of over \$521 million related to such scams. To complement this analysis, we performed a survey on crowdsourcing platforms with 584 users to broaden the insights on comparative analysis of pig-butchering scams with other types of scams. Our research highlights that these attacks are sophisticated and often require multiple entities, including policymakers and law enforcement, to work together alongside user education to create a proactive detection of such scams. 

\end{abstract}
\section{Introduction}

According to the Federal Bureau of Investigation (FBI), in 2022, investment-related fraud resulted in \$3.31 billion in losses~\cite{binancerise23,techtarget24}. 
In 2023, this type of fraud accounted for \$4.57 billion, an increase of 38\% over the previous year~\cite{nco24}. In 2024 alone, the FBI received 18K complaints, reporting \$1.9 billion losses~\cite{abc7}. These loss metrics are only accounted from reported ones and many go unreported as victims do not report being scammed due to several psychological, emotional, and social factors~\cite{latimesShame,nationaTradingShame,wtkrShame}.

In recent years, a specific type of investment-related fraud became prominent: the phenomenon of \emph{pig-butchering} scams has emerged as a significant threat in the landscape of social engineering~\cite{ciscotalos24,coindesk24,aurora24,proofpoint24}. The term \emph{pig-butchering} is derived from the Chinese phrase \emph{Sha Zhu Pan}, where fraudsters establish trust with a victim through romance or a similar trustworthy relationship, metaphorically ``fattening the pig'' before conning them~\cite{investopedia24}. Fraudsters later deceive the potential victim into investing via a fake investment platform or asking for a transfer of funds making a fake emergency support before finally ``butchering'' them. Overall, pig-butchering scams are a more recent and sophisticated evolution of romance~\cite{romancescamfbi,romancescamFTC} and investment scams~\cite{investmentscamFTC,investmentscamDPFI}, where fraudsters exploit the popularity and complexity of cryptocurrency to deceive victims. These scams involve emotional manipulation or promises of quick, low-risk returns through fraudulent investments.

Pig-butchering scams typically begin with fraudsters reaching out to potential victims through social media~\cite{socialmediastart,mnPartners24,forbes24}, dating apps~\cite{datingappstart,newyorkerdatingtimes,nltimesdating}, or other online platforms~\cite{tenable24,tenableotherplatforms24}. Such scams are often orchestrated by gangs of scammers, or by abducted and trafficked humans who are forced to perform scams on target their victims, establishing relationships that last weeks to months~\cite{wiredforcedhuman,humantrifficlinkedIn}. In a fraudulent investment scenario, the victim receives some profitable returns upon investing and can withdraw, which builds the credibility of the investment whereas for romance fraud, the victim is asked to support the fraudsters in emergency financial support before the victim, and fraudsters can physically meet. In both cases, when the victim either increases their transfers or investments to larger amounts or can no longer transfer or invest, the fraudulent investment platforms block withdrawals, citing fake technical issues, or they go offline altogether. In romance scams, messaging apps, social media, or dating profiles eventually go offline as well, leaving the victim defrauded~\cite {hackernews24}.

Although the security community has recently begun exploring pig-butchering scams via technical reports~\cite{evercWhitePaper,cloudSEKWhitePpaer,trendMicroWhitePpaer,federalResearchWhitePpaer}, and academic papers~\cite{wang2024victim, jiang2023emotional, maras2024deconstructing, wang2023schemes, whitty2015anatomy, tan2023network,bilz2023tainted}, there is still a lack of detailed understanding of how fraudsters set up social media profiles and engage with victims using various social engineering techniques as part of the scam orchestration. While the prior research has examined various aspects of financial fraud and cybercrime, including phishing, identity theft, and investment fraud, the specific phenomenon of pig-butchering scams remains under-explored. Existing literature often focuses on the economic impact of such crimes and general strategies for prevention and detection~\cite{anderson2019financial,anggusti2022cybercrime,chawki2022cybercrime}, whereas exploration of scammer engagement including multiple social media platforms and channels of communication was not studied yet.

In this paper, we systematically study fraudsters carrying out pig-butchering scams from three sources: (i) social media platforms such as \emph{X} (formerly known as Twitter), \emph{Instagram}, \emph{Telegram}, and \emph{YouTube}, (ii) public reported abuse databases such as \emph{Chainabuse}~\cite{chainabuse}, and \emph{Crypto Scam Tracker}~\cite{crytposcamtracker}, and (iii) news articles case studies on pig-butchering scams. Through these three vantage points, we collect the victim's direct or attempted pig-butchering experiences and provide a detailed analysis of the scam life cycle. Additionally, we perform quantitative studies through crowd-sourced surveys to further add detail to online scams and fraudsters' strategies.

More specifically, we performed the first large-scale study of pig-butchering via (i) multiple social media platforms, collecting 431,731 social media accounts and 771,245 posts, with 146 confirmed victims of pig-butchering scams; (ii) collecting 3,213 narratives related to abuse report on public database with 2,570 confirmed narratives of being a victim of pig-butchering scams; (iii) collecting 1,074 news outlets, through automated and qualitative analysis confirming 50 unique case studies related to 834 victims of pig-butchering. We performed tracking and evaluation of scam mechanics including scammer's fraudulent channels and payment method. Through this study, we approximated the total loss from victims (146 social media users, 2,570 abuse databases narratives, and 50 case studies of 834 souls) collectively losing over \$521 million tied to pig-butchering scams. Additionally, we performed a quantitative study via a crowd-source platform with 586 participants to broader understand the online scams experience of in-the-wild users, and provide a comparative analysis with pig-butchering scams. Finally, we provide recommendations to better defend against such scams in the future. 

\smallskip
\noindent We summarize our key contributions as follows:

\begin{itemize}
    \item \textbf{Large Scale Study on Pig-butchering Scam.} We present the first large-scale study on pig-butchering scams across three sources: social media platforms, abuse-reported databases, and new articles on first-hand reports of experiences with fraudsters operating globally. Additionally, we perform user studies (n=586) via a survey to provide further insights on online scams in comparison to pig-butchering scams. 
    
    \item \textbf{Scam Mechanics and Fraud Tracking.} We provide a comprehensive analysis of the modus operandi of scammers executing pig-butchering schemes, identifying their fraudulent schemes and the payment methods they use as part of these scams. Our research provides an in-depth analysis of fraudsters' life cycle operations that are orchestrated via various platforms.
    
\end{itemize}
   
To foster research, we share our code~\cite{pigbutcheringcode} and data related to victim's experiences. However, for data protection reasons, the data related to identifying scammers (e.g., social media profiles, emails, URLs, and cryptocurrency addresses) are only shared with interested academics, abused entities, or researchers upon request.

\section{Related Work}

To the best of our knowledge, this work is the first large-scale, systematic study to conduct a comprehensive analysis of pig-butchering scams using several data sources.  
Below, we discuss relevant prior studies, highlight the unique aspects of our research, and address the existing research gap.

\BfPara{Cryptocurrency, Investment, and Romance Scams} Previous studies have investigated various types of abuse and scams in cryptocurrency~\cite{9591634,9493255,li2023nothing,huang2018tracking}, investment~\cite{siu2022,gao2020,kuo2024constructing}, and romance domains~\cite{buchanan14,whitty12,whitty2018you,cross2024romance}. For instance, Bartoletti et al.~\cite{9591634} examined the prevalence of cryptocurrency scams and developed a taxonomy of the types of attacks used by cryptocurrency fraudsters. Xia et al.~\cite{9493255} focused on cryptocurrency scams that emerged during the COVID-19 pandemic. In the area of romance fraud, Buchanan et al.~\cite{buchanan14} and Whitty et al.~\cite{whitty12} analyzed online romance scams where fraudsters target potential victims by pretending to seek an intimate relationship. 
However, none of the previous studies performed a comprehensive end-to-end tracking of cryptocurrency and the life-cycle of pit-butchering scams in comparable contexts.

\BfPara{Pig-butchering Scams} Although pig-butchering scams are relatively recent, researchers from various fields have begun examining the abuse~\cite{burton2024pig,cross24,maras2024} and its impact on victims~\cite{wang2024victim,jiang2023emotional,tan2023network}. For example, Wang~\cite{wang2024victim} describes the experiences of trafficked Chinese workers who are forced into pig-butchering scams. Burton et al.~\cite{burton2024pig} provides an overview of the methodologies behind these scams through a literature review survey, and Cross et al.~\cite{cross24} analyze the evolution of social engineering tactics used by fraudsters in romance and cryptocurrency scams. The close work from ours by Maras et al.~\cite{maras2024} focuses on investment fraud, analyzing news articles and court documents with an emphasis on criminal justice practices. 

\BfPara{Abuse study on Social Media Platforms} Over the past five years, social media has become a key platform for studying scams and abuses across various topics, including 
cryptocurrency scams~\cite{acharya2024conning,chandra2024detection}, brand and user attacks~\cite{roberts2019you,acharya2024imitation}, hate speech~\cite{rawat2024hate,tabassum2024investigating,arunasalam2024understanding}, 
and psychological abuse~\cite{franco2024characterizing,alshamrani2020detecting}.
With \emph{HoneyTweet}~\cite{acharya2024conning}, Acharya et al. examined fake technical support scams targeting popular cryptocurrency wallet users, while Ratkiewicz et al.~\cite{ratkiewicz2011detecting} investigated the tracking and detection of political abuses propagated through social media. 
Despite these efforts, a research gap remains in identifying first-hand victims involving online platforms in pig-butchering scams.

\BfPara{Study on Public Abuse Reports} Previous studies on public abuse reports have examined various aspects of cryptocurrency-related abuse, such as categorizing types of cryptocurrency abuse~\cite{gomez2024sorting,choi2022large,klein2024fighting}, tracking abuse campaigns on the dark web~\cite{xia2024devil,topor2019cyber}, and analyzing infrastructure models in abuse reports~\cite{jhaveri17,parti2023if}. However, no prior research has focused specifically on abuse reports of victims of pig-butchering scams.

\section{Evaluation Setup and Methodology}
\label{sec:evaluation_set_up}
\begin{figure}[tb]
\centering
\includegraphics[width=.43\textwidth]{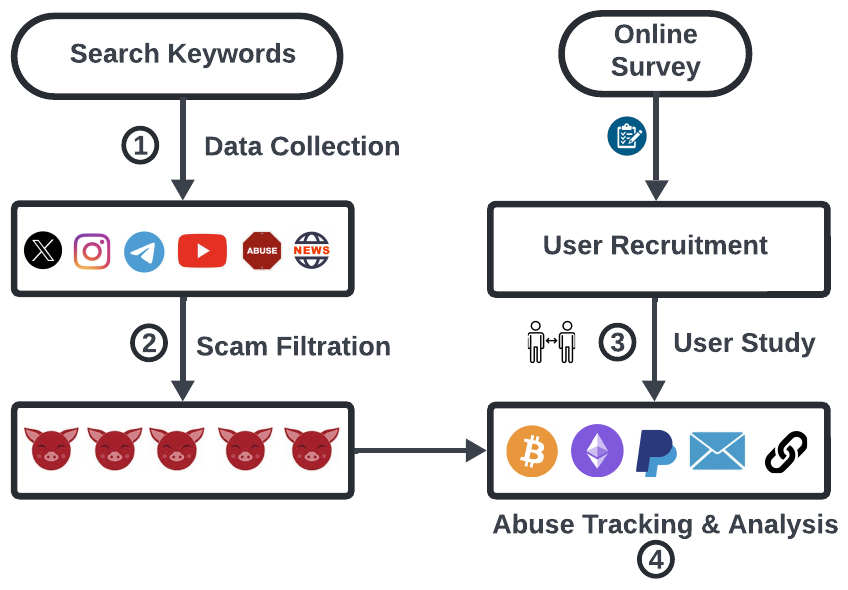}\hfill
\caption{Data Collection – Our system comprises three main modules. Based on our manual observations of direct victims of pig-butchering scams, we developed search keywords to collect data from various sources (\circleone). After collecting the data, we apply automated and manual filtration of data (\circletwo). Additionally, we perform quantitative studies via survey on online scams (\circlethree), and finally, we evaluate the collected data by tracking and profiling abusive elements and victims' narratives (\circlefour).}
\label{fig:sys_design}
\end{figure}

In this section, we present our evaluation setup and methodology to understand the life cycle of a pig-butchering scam. Our system is composed of three main modules, as illustrated in \autoref{fig:sys_design}: \circleone gathers data from three sources: (i) social media platforms, (ii) publicly reported abuse incidents related to pig-butchering scams, and (iii) news articles related to pig-butchering scams collected from multiple search engines; \circletwo performs semi-automated filtration on such collected data to ensure the data are related to this kind of scams; \circlethree performs the quantitative study on users experiencing online scams in the last five years recruiting through a crowd-sourced platform, and \circlefour analyzes the aggregated data collected to further validating the fraudulent activities via tracking scamming profiles and abusing payment methods. We provide descriptions of our approach below and discuss ethical considerations and the disclosure process in Appendix~\ref{sec:discussion}.

\subsection{Search Methodology and Raw Dataset}
\label{sec:raw_dataset}
From March 2024 to August 2024, we collected data from three primary sources: (i) social media platforms, (ii) public reports submitted to cryptocurrency abuse databases, and (iii) news articles about pig-butchering scams. For organizing relevant data searches on social media, we manually crafted keywords based on observations of public posts across various platforms. For abuse reports and news articles, we focused keyword searches on the terms ``pig-butchering'', ``romance scam'', and ``investment scam''. Below, we provide further details on our search methodology and raw data collection process. We discuss the potential limitations of the data sets in Appendix~\ref{sec:discussion}.

\subsubsection{Manual Search Keywords Collection} During our research incubation period, we performed a manual search on multiple social media platforms to identify cases and abusive profiles related to pig-butchering. From our observations, we identify three categories of pig-butchering schemes targeted to individuals looking for an \emph{online dating/romance}, \emph{investment}, and \emph{job}. We observe first-hand victims mentioning these stories throughout their posts. We take these three schemes as ground truth for the formation of keywords in social media posts to search for relevant pig-butchering posts. Based on our manual observation, we created a total of 219 keywords that we used as part of search posts on social media platforms. In Appendix~\ref{sec:manual_keywords}, we provide further detail on this keyword-gathering methodology. For abuse databases, we focused our data search on the scam tag classification provided by abuse databases from \emph{Chainabuse} and \emph{Crypto Scam Tracker}. In reviewing these databases, we found that numerous complaints were categorized as pig-butchering scams, investment fraud, or romance scams. We included investment fraud and romance scams in our analysis because pig-butchering scams often involve advanced social engineering tactics, where victims are groomed over a prolonged period. As a result, we focused our analysis on reports marked with these classifications.  

\subsubsection{Semi-Automated Data Collection} 
We collected data from social media platforms by performing automated API calls using search keywords, while data from abuse databases was gathered through a combination of collaboration and manual downloads. We provide details on each source below. 

\BfPara{Social Media Data} We perform keyword-based searches on four social media platforms: \emph{X}, \emph{Instagram}, \emph{Facebook}, and \emph{Telegram}. In particular, we automated API services~\cite{TwitterUserDetailAPI,TwitterTimelinesAPI,InstagramScraper,TelegramApifyScraper,TelegramTelemetrScraper,YouTubeScraper} to collect the data from these social media platforms. We provide a breakdown of the raw dataset from each of these social media in~\autoref{table:social_media_raw_dataset}. In total, we collected 789,751 posts from 432,762 users from these platforms. For each of the social media users, we further collected the profile metadata such as profile name, descriptions, location, followers, and profile image. Among this platform, \emph{X} comprised the highest number of posts and accounts---overall 53\% (414,992/771,245) of the posts comprised 76\% (328,822/431,731) of accounts stored in our database. The lowest count resulted from \emph{YouTube}, 21\% (93,618/431,731) accounts comprised of 9\% (71,607/771,245) posts. Our dataset from all four social media platforms had a median of 49,182 accounts and 142,323 posts.

\begin{table}[tb]
    \small
    \caption{Overview of the raw dataset obtained by performing search queries across four social media platforms. Among the four social media platforms, we observe that \emph{X} contains the largest number of accounts and posts.}
    \centering
    \begin{tabular}{lrrr}
        \toprule
        \rowcolor{gray!0}
        \multicolumn{1}{c}{\textbf{Platform}} & \multicolumn{1}{c}{\textbf{Accounts}} & \multicolumn{1}{c}{\textbf{Distinct Posts}}  & \multicolumn{1}{c}{\textbf{All Posts}}\\
        \midrule
		X & 328,822 & 125,264 & 414,992 \\
            \rowcolor{gray!10}
		Instagram & 4,746 & 175,000 & 190,236 \\
            \rowcolor{gray!0}
		Telegram & 4,545 & 94,410 & 94,410 \\
            \rowcolor{gray!10}
		YouTube & 93,618 & 65,295 & 71,607 \\
            \rowcolor{gray!0}
        \midrule
		All & 431,731 & 459,969, & 771,245 \\
            \rowcolor{gray!0}
		\bottomrule
    \end{tabular}   
    \label{table:social_media_raw_dataset}
\end{table}

\begin{table}[tb]
    \small
    \caption{Overview of the raw dataset on news articles related to pig-butchering scams obtained by performing search queries across three search engines via web search and news search methodology. Among the three search engines, we observe that \emph{Bing} contains the largest number of articles.}
    \centering
    \begin{tabular}{lrrr}
        \toprule
        \rowcolor{gray!0}
        \multicolumn{1}{c}{\textbf{Search Engines}} & \multicolumn{1}{c}{\textbf{Web Search}} & \multicolumn{1}{c}{\textbf{News}}  & \multicolumn{1}{c}{\textbf{All Articles}}\\
        \midrule
		Yahoo & 167 & 110 & 172 \\
            \rowcolor{gray!10}
            Google & 157 & 355 & 477 \\
            \rowcolor{gray!0}
            Bing & 237 & 466 & 682 \\
            \rowcolor{gray!10}
        \midrule
		All (Distinct) & 355 & 871 & 1074 \\
            \rowcolor{gray!0}
		\bottomrule
    \end{tabular}   
    \label{table:news_article_dataset}
\end{table}
    
\BfPara{Public User Reported Data} We collected public user reports on pig-butchering scams, particularly those involving narratives tagged with pig-butchering scams. We collaborated with \emph{Chainabuse}, a well-known cryptocurrency abuse reporting platform, for fetching the data associated with public reports on pig-butchering. The second data source was manually downloaded from~\emph{Crypto Scam Tracker}, Department of Financial Protection and Innovation, Official website of the State of California. In~\autoref{table:abuse_db_reports}, we provide a summary of the public reports gathered from both sources. In total, we collected 3,213 public narratives associated with 1,710 distinct complaints. 

\begin{table}[tb]
    \small
    \caption{Overview of the abuse database: In this table, we highlight pig-butchering scam complaints gathered from two public sources. These public reports include victim complaints about close encounters with scammers, along with narrative details describing their interactions.}
    \centering
    \begin{tabular}{lrr}
        \toprule
        \rowcolor{gray!0}
        \textbf{Abuse DB} & \textbf{Reports} & \textbf{Narratives} \\
        \midrule
        Chainabuse & 1,467 & 2,970 \\
        \rowcolor{gray!10}
        Crypto Scam Tracker & 243 & 243 \\
        \rowcolor{gray!0}
        \midrule
        Total & 1,710 & 3,213 \\
        \rowcolor{gray!10}
        \bottomrule
    \end{tabular}    
    \label{table:abuse_db_reports}
\end{table}

\BfPara{Public News Articles} We collected publicly available news articles on pig-butchering scams using a custom Python Selenium automation. For this process, we utilized three search engines: \emph{Google}, \emph{Bing}, and \emph{Yahoo}. Our choice of these search engines is motivated by their popularity~\cite {topsearchengines}. For each search engine, we automated searches across their \emph{web search} and \emph{news search} features using the keyword \emph{pig-butchering scam}. For the search limits, we restricted the web search to the first 30 pages, while the news search was limited to the maximum scroll permitted by each search engine on its news page. As summarized in \autoref{table:news_article_dataset}, we collected a total of 1,074 news articles from these sources, with \emph{Yahoo} contributing the fewest at 16\% (172/1,074) and \emph{Bing} the most at 63\% (682/1,074). 

\subsection{Data Filtration and Candidate Selection}
\label{sec:data_set_filtration}
We applied three distinct data filtration processes to the collected data. First, for social media platforms, we used prompt queries backed by large language models (LLMs) to identify posts from direct victims of pig-butchering scams. Second, for public reports, we ensured that each report was consistently tagged by users as a pig-butchering scam complaint and conducted manual reviews of each narrative. Lastly, for news articles, we filtered out irrelevant articles and performed a thorough qualitative review. We provide additional details on each of these sources as below.

\subsubsection{Data Filtration on Social Media}

To identify victims of pig-butchering scams, we initially applied two automated methods: (i) keyword heuristics to assess post engagement, specifically targeting users who mentioned being scammed by referencing terms like \emph{scam}, \emph{fraud}, and \emph{lost}, along with first-person pronouns (\emph{I}, \emph{me}, and \emph{my}); and (ii) prompt engineering with large language models (LLMs) to detect content related to specific fraudulent activities, including pig-butchering, romance scams, investment scams, and cryptocurrency scams. Through these automated approaches, we identified 0.27\% (2,096/771,245) of distinct user narratives as potential scam reports from the overall raw dataset, sharing 123 narratives from both automated techniques: LLMs, and keyword heuristics. Acknowledging prior work that LLMs can produce hallucinations~\cite{llmhallicunateFlorian}, and Natural Language Processing (NLP) heuristics could contain inaccurate content filtering~\cite{khurana2023natural}, we then conducted a manual dataset evaluation of these posts across four social media platforms, confirming 146 accounts sharing their personal experience of pig-butchering scams. During the manual review, we ensured that identified pig-butchering cases involved extended grooming periods, distinguishing them from standard romance and investment scams. In~\autoref{table:social_media_victim_dataset}, we present a breakdown of filtered victims by social media platform and methodology, with additional details on prompt-engineering filtration available in the Appendix \ref{sec:social_media_data_filtration}.

\begin{table}[tb]
    \small
    \caption{Overview of social media-based victim filtration from the raw dataset applying automated (LLMs, and NLP-heuristics) and manual reviews. Between the two filtration methods, we found that LLM-based filtering identified more victims than NLP-based heuristics. Additionally, \emph{YouTube} had the highest number of users willing to openly share their experiences as victims compared to other social media platforms.}
    \centering
    \begin{tabular}{lrrr}
        \toprule
        \rowcolor{gray!0}
        \multicolumn{1}{c}{\textbf{Platform}} & \multicolumn{1}{c}{\textbf{LLMs}} & \multicolumn{1}{c}{\textbf{NLP-Heuristics}}  & \multicolumn{1}{c}{\textbf{Manual Reviews}}\\
        \midrule
		Instagram & 2/84 & 7/372 & 9 \\
            \rowcolor{gray!10}
		  Telegram & 10/228 & 2/151 & 12 \\
            \rowcolor{gray!0}
		  X & 5/69 & 29/321 & 34 \\
            \rowcolor{gray!10}
		YouTube & 68/221 &  23/835 & 91 \\
            \rowcolor{gray!0}
        \midrule
		All & 85/602 & 61/1679 & 146 \\
            \rowcolor{gray!0}
		\bottomrule
    \end{tabular}   
    \label{table:social_media_victim_dataset}
\end{table}

\subsubsection{Data Filtration on AbuseDB} 

We conducted a manual review of 3,213 narratives from 1,710 posts to ensure our collected data accurately represented cases specific to pig-butchering scams. In our manual review, we classified narratives based on two scenarios: (i) cases tagged as pig-butchering that contained narratives clearly showing characteristics of pig-butchering scams, and (ii) cases tagged under related categories, such as romance and investment fraud, that involved prolonged victim grooming were reclassified as pig-butchering. We included specifically romance and investment fraud cases because pig-butchering scams are often structured around similar tactics of prolonged victim grooming. During this review, we filtered out duplicated reports, narratives related to other types of crypto scams (such as sextortion or blackmail), and entries with insufficient information to confirm relevance to pig-butchering scams. Through this filtering process, we excluded 13\% (232/1,710) of the reports and 19\% (642/3,213) of the narratives. As a result, our final dataset of abuse reports includes 1,478 unique reports comprising 2,570 narratives.

\subsubsection{Data Filtration on News Articles} Our news filtration process involves several semi-automated steps. First, we filtered out 97 URLs linked to unrelated content, such as \emph{YouTube} videos and social media posts from \emph{X} or \emph{Instagram} about pig-butchering. For the remaining 977 URLs, we performed a Python URL alive check to confirm active links, retaining only those with response codes between 200 and 300. This process filtered out an additional 19 inaccessible URLs. To ensure the content was relevant to pig-butchering, we developed a custom Selenium Python script to retrieve page content, identifying 410 URLs containing pig-butchering context. Recognizing that some automated page visits are blocked, we checked for \emph{CAPTCHA} terms to be present on the page source content, identifying 92 URLs restricted by \emph{CAPTCHA}. From the multiple stages of filtration, we selected 501 URLs as candidate links and performed a qualitative manual review of them.

\subsection{User Study}
\label{sec:user_study}
Our third module conducts a representative quantitative study on users' experiences with online scams over the past five years. For this study, we recruited survey participants from crowd-sourcing platforms to assess the frequency of online scams and identify which types are most common, including classic scams such as phishing, technical support fraud, online shopping scams, and identity theft. Additionally, we provide a comparative analysis of users' experiences with pig-butchering scams. Through this study, our aim was to gain a deeper understanding of victim experiences, particularly with pig-butchering scams, and to provide broader insights into various other types of scams.

\subsection{Tracking and Analysis}
\label{sec:tracking_and_analysis}
The fourth module, \emph{abuse tracking and analysis}, offers insights into scammer engagement, along with quantitative analysis of victim losses, impacts, and the methods fraudsters use to lure victims. This module performs an in-depth analysis of various elements associated with both pig-butchering scams and their victims. It includes features such as engaged posts, victim narratives, scam-related social media profiles, linked payment methods cryptocurrency addresses, and operational techniques used throughout the scam’s life-cycle. 

\medskip
\textbf{Paper Outline.}
For the rest of the section organization, we present our findings as follows: social media abuse measurement in~\autoref{sec:social_media_prevalence}; public reported abuse reports evaluation in~\autoref{sec:public_abuse_report}; evaluation of new media and coverage on victims of pig-butchering in~\autoref{sec:news_article_evaluation}; tracking fraudulent communication channels, external URLs, and payment methods of scammers in~\autoref{sec:financial_tracking_and_analysis}; representative quantitative users study experiencing online scams in~\autoref{sec:quantitative_user_study} and ethical consideration and data limitations in~\autoref{sec:discussion}. Summarizing our findings, and insights collected from the victim's experiences, we provide the recommendations in tackling pig-butchering scams in~\autoref{sec:recommendations}.

\section{Social Media Abuse Measurement}
\label{sec:social_media_prevalence}

In this section, we provide the qualitative analysis of 146 pig-butchering victims found on social media, focusing on three key areas: (i) confirmation of financial losses, (ii) the scammer’s methods and tactics used in the operation, and (iii) the victim’s post-scam experiences, including the impact on their lives as described in their public posts. 

\begin{figure}[tb]
\centering
\includegraphics[width=.45\textwidth]{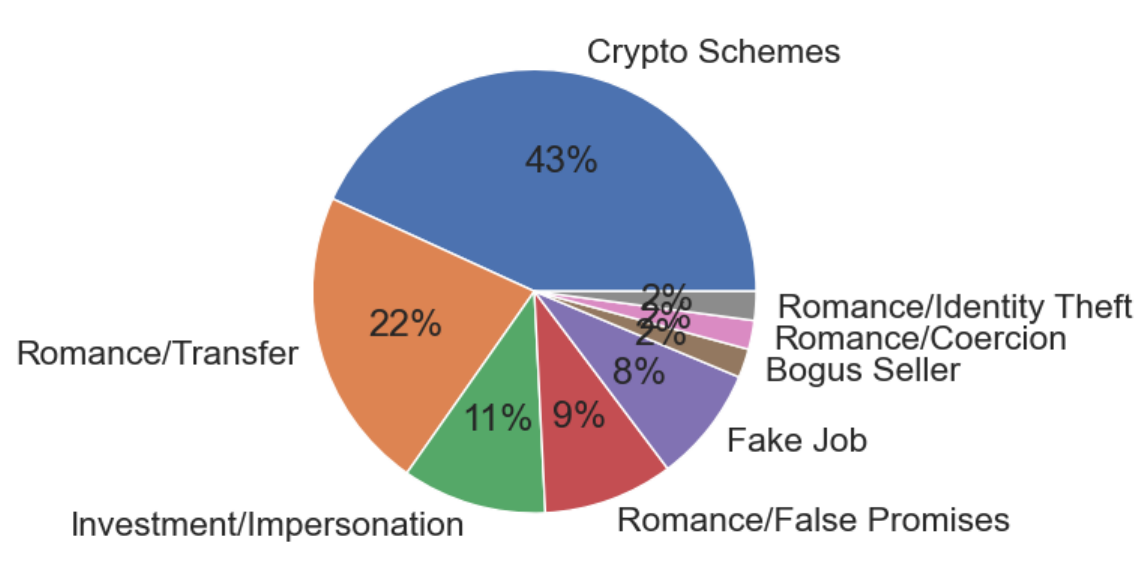} \hfill
\caption{Scam Techniques in Victim Reports on Social Media Platforms - The graph presents eight scam techniques used by scammers in social engineering tactics associated with pig-butchering scams. Our findings indicate that \emph{Crypto Schemes} are the most prevalent, accounting for 43\% of all victim reports, while Romance scams involving \emph{Identity} and \emph{Coercion} are among the least reported.}
\label{fig:social_media_scam_tactics}
\end{figure}

\textbf{Overview.} In \autoref{table:social_media_victim_amount_loss}, we show the total reported losses by victims on each social media platform. Of the four platforms, victims on \emph{YouTube} reported the highest total loss, with \$14,341,820 from 58/91 victims, while \emph{Instagram} showed the lowest reported loss of \$2,200 from 2/9 victims. Our analysis reveals that 57\% (84/146) of victims openly disclosed their financial losses due to scams. Additionally, 65\% (95/146) of victims shared details on the specific social engineering techniques used by scammers. We identified eight distinct scamming tactics: Crypto Schemes (41/146), Romance scams involving financial transfers (21/146), Investment/Impersonation (10/146), Romance scams with false crypto investment promises (9/146), Fake Job offers (8/146), Bogus Seller Business Setup (2/146), Romance scams leading to online coercion (2/146), and Romance scams resulting in Identity Theft (2/146). In \autoref{fig:social_media_scam_tactics}, we illustrate the breakdown of these scam techniques based on victim reports on each social media platform. We provide additional details on each social media platform below.

\begin{table}[tb]
    \small
    \caption{Approximated Dollar amount losses from the disclosed victim from social media dataset - This table provides an estimated dollar value for losses reported by victims. It includes the lower-bound approximated financial losses, with international currencies and cryptocurrencies converted to USD based on exchange rates and cryptocurrency values from the first week of November 2024.}
    \centering
    \begin{tabular}{lrrr}
        \toprule
        \rowcolor{gray!0}
        \multicolumn{1}{c}{\textbf{Platform}} & \multicolumn{1}{c}{\textbf{Disclosed}} & \multicolumn{1}{c}{\textbf{All Victims}}  & \multicolumn{1}{c}{\textbf{Approx. Amount}}\\
        \midrule
		Instagram &  2 & 9 & \$2,200 \\
            \rowcolor{gray!10}
		  Telegram & 6 & 12 & \$371,538 \\
            \rowcolor{gray!0}
		  X & 18 & 34 & \$426,745 \\
            \rowcolor{gray!10}
		YouTube & 58 & 91 & \$14,341,820 \\
            \rowcolor{gray!0}
        \midrule
		All & 84 & 146 & \$15,142,303 \\
            \rowcolor{gray!0}
		\bottomrule
    \end{tabular}   
    \label{table:social_media_victim_amount_loss}
\end{table}

\subsection{Evaluation on Instagram} Among the four social media platforms, we identified 6\% (9/146), the lost number of victims sharing their experiences that relate to pig-butchering scams. We provide further details below. 

\BfPara{Victim Confirmation and Financial Looses} As Instagram is widely used for sharing photos and videos, we suspect that victims are reluctant to share posts compared to other social media platforms. Among the 9 victims' experiences, only two victims shared their experience of losing between \$200 -- \$2000 in package delivery scams, where fraudsters lured victims into investing in “unclaimed package” boxes and such boxes never arrived to victims. 

\BfPara{Scam Tactics} We identified four distinct natures of scam tactics as part of a shared experience of being scammed. These include (i) 4/9 fraudulent crypto schemes (e.g., \emph{OneCoin}, \emph{EXW Wallet}), (i) 4/9 romance scams involving financial requests (iii) 2/9 bogus seller scams (e.g., premium Netflix accounts, unclaimed package sales), and (iv) 1/9 in-person impersonation scams scammer posing as official personnel, visiting victim in person with counterfeit ID. 

\BfPara{Victim's Emotional Experience} Out of the nine victims, two reported receiving a ring from a scammer, only to later realize it was a scam. Another two victims filed multiple complaints with the Better Business Bureau regarding the unclaimed package investment scheme, expressing frustration that their complaints were ignored, resulting in unresolved financial losses.

\subsection{Evaluation on Telegram}
Victims on Telegram account for 8\% (12/146) of our overall dataset, making it the second-lowest platform by victim count, following Instagram. Our evaluation of victims' narratives on Telegram is summarized below.

\BfPara{Victim Confirmation and Financial Losses} Of the 12 victims identified on Telegram, 6 victims disclosed their financial losses, while 6 did not reveal the amount lost. Among those who reported their losses, 3 victims collectively lost a total of 110 ETH, with individual losses of 10 ETH, 40 ETH, and 60 ETH. Additionally, two victims reported losing 425 INR, 175 INR, and 250 INR, respectively. The remaining victim lost 5M \$BLV tokens, although the exact value of the loss is unclear. The total amount approximated to \$371,538 based on the currency conversation rate of the first week of November 2024.

\BfPara{Scam Tactics} We identified three distinct types of scam tactics related to employment and investment scams. These include: (i) 7/12 victims experienced stolen funds and tools, where several victims reported losing large amounts of cryptocurrency, often during private sales or pre-sale events, which they later discovered to be fraudulent; (ii) 3/12 victims were affected by escrow-related fraud, where a deal was arranged through an escrow system, the victim agreed to work, but the scammer failed to make payment; and (iii) 2/12 victims were misled by false promises of refunds or future compensation, with scammers assuring refunds once liquidity pools were unlocked.

\BfPara{Victim's Emotional Experience} We observe three main emotional experiences shared by 9/12 victims. These include (i) (4/9) messages warning others about specific scammer naming them and their experiences to public Telegram forums, (ii) (2/9) acceptance and moving, reflecting on the loss with a degree of acceptance, stating that it’s better to recover something rather than nothing. These individuals also emphasize the importance of moving forward and not dwelling on the scam, and (iii) (3/9) emotions like anger, frustration, and even suicidal thoughts are mentioned, where one of the victims shared mentions, “Before I die, I will make sure I kill this scammer" an indication of the serious emotional toll these scams take on victims.

\subsection{Evaluation on X}
While \emph{X} platform had the highest number of posts and accounts in our raw dataset, it ranks second in reported victim count, representing 23\% (34/146) of the total. Below, we provide victims' shared experiences and highlight the details collected from our analysis.  

\BfPara{Victim Confirmation and Financial Looses} Out of 34 individuals, 32 individuals shared their victim's experience with confirmed financial losses, while 2 narrowly avoided being scammed by growing suspicious during their interaction. Of the 32 confirmed victims, 14 shared their emotional experiences without specifying their actual financial losses, while 18 reported both the emotional impact and the amount lost. Among those who disclosed amounts, losses ranged from over \$250,000—the highest reported, tied to a life-saving investment fraud—down to \$200, which was requested under the pretext of a romantic gesture. The median loss among victims was \$2,500. From these, two reported their losses in Ethereum, and two in Euros. We approximated the dollar values for Ethereum and Euros. In total, we approximated, \$426,745 based on the conversion rates of the first week of November 2024.

\BfPara{Grooming Period} Regarding the duration of grooming, 21 cases did not specify a timeline, 9 described scams occurring within a short span (1-6 weeks), and 7 mentioned prolonged periods exceeding 7 weeks. One case reported a scam that lasted over three years as an intermittent relationship.

\begin{figure}[tb]
\centering
\includegraphics[width=.43\textwidth]{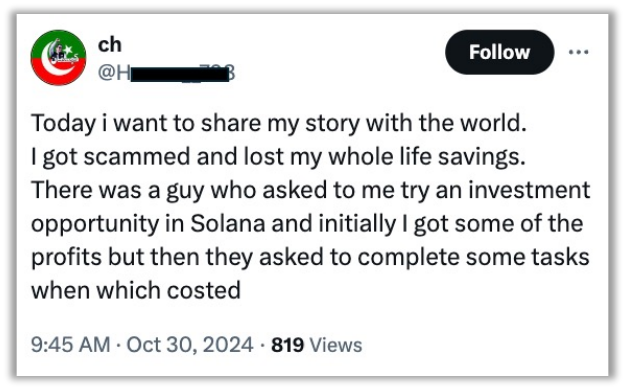}\hfill
\caption{In this figure, we display \emph{X} posts from a user describing the experience of a fraudulent cryptocurrency investment-based pig-butchering scam. The victim shares an experience of being scammed to raise awareness about fraudulent investment schemes.}
\label{fig:victim_story}
\end{figure}

\BfPara{Scam Tactics} We identified 25/34 victims who reported scammer's tactics as part of being scammed. These include romance scams with false promises (9 cases), investment fraud through fake tokens (4) (see~\autoref{fig:victim_story}), high-profile impersonation in investment fraud (4), celebrity impersonation in romantic schemes (4), online coercion in romantic scams (2), and identity theft in romance scams (2).

\BfPara{Victim's Emotional Experience} We observe that 20/25 victims' narratives share their feelings of shame, self-blame, and distrust following scams, showing that these scams impact mental health as well as finances. Examples of shared emotional impacts include (i) (6/25) the victim’s accounts being blocked, left in emotional distress, (ii) (4/25) the victim being left with debt, unaware until receiving a default notice or bills, (iii) (3/25) led to life-altering decisions, (iv) (3/25) victim feels devastated, looking for community support, (v) (2/25) victim suffered emotional trauma, blames self, leads to lasting shame, and (vi) (2/25) perpetrator used victim’s photos to scam others, leading to guilt or shocked.

\subsection{Evaluation YouTube}

Across the four social media platforms, we identified 62\% (91/146) of the total shared victim experiences, the highest proportion among them. We present our findings in four key areas, with insights detailed below.

\BfPara{Victim Confirmation and Financial Looses} For YouTube, we relied on YouTube's description part of the YouTube channel in gaining the experience shared by the victim. We only included descriptions that provided the direct experience related to the individuals rather than a generic channel of pig-butchering. Out of 91 victims' experiences, we identified three different confirmations: (i) 82/91 channels featured on behalf of victims sharing their losses, and emotional experiences, (ii) 5/91 self-featured channels by victims, and (iii) 4/91 shared nearly being scammed by scammer, and raising awareness throughout the video sharing their tips on the ongoing the pig-butchering scams. Among the nationality-shared information, we identified 58/91 victims from 7 different nationals: US (27), Singapore (9), United Kingdom (8), India (6), Canada (4), New Zealand (2), and Australia (2). The loss amount reported total from eight currencies: USD (4,419,000), SGD (99,000), EUR (3,977,100), GBP (3,314,250), INR (362,358,000), CAD (500,000), AUD (850,000) and JPY (574,470,000), totaling to \$14,341,820 based on the currency conversion rate of fist week of November 2024.

\BfPara{Grooming Period} Our analysis of victim reports revealed varying grooming periods based on the type and complexity of scams: (i) cryptocurrency investment scams typically lasted 3-8 weeks, (ii) romance scams involving a relationship spaned  4 weeks to 6 months, (iii) romance scams with an investment took 2-4 months, (iv) romance scams with fake celebrities were discovered by victims with suspicious behavior within hours to days, and (v) long-term manipulation in romance scams, where scammers repeatedly extract money, reported to last several months to a year. These cases highlight that pig-butchering scams involving financial manipulation, especially romance scams, scammers engage in longer grooming to build trust with victims to facilitate significant financial transactions.

\BfPara{Scam Tactics} From the shared experience, we observed scammers performing various social engineering tricks on users: (i) 24/91 victims resulted in high-value losses from \$100K to over \$1Mil from fraudulent cryptocurrency investment, (ii) 17/91 includes cases like fake romantic partners asking for money, (iii) 5/91 scammers performing impersonation with fake identity use posing as military, doctors, or wealthy investors, and 3/91 scammer performed job fraud / fake opportunities that result in money losses. 

\BfPara{Victim's Emotional Experience} We observe 38/91 victims' complex emotional experiences being shared, often affecting multiple aspects of their lives and relationships. These include: (i) the victim felt a sense of betrayal after deeply getting connected with the scammer (12/91), (ii) faced significant losses leading to despair over life's savings (8/91), (iii) life after the scam impacted the family dynamics and relationship (5/91), (iv) anxious and fearful about financial security and suicidal thoughts (4/91), (v) anger and frustration at the difficult to recover (3/91), (vi) embarrassed to share and self low esteem (3/91), (vii) despite the trauma, 3/19 victims felt need to share their stories to the media. 
\section{Public Abuse Reports Evaluation}
\label{sec:public_abuse_report}
In this section, we present an evaluation of 2,570 victim narratives gathered from two sources: \emph{Chainabuse} and \emph{Crypto Scam Tracker}. Our analysis focuses on five key categories: (i) the initial contact method used by scammers, (ii) how scammers build relationships with victims, (iii) the techniques scammers employ to steal funds, (iv) emotional and psychological manipulation, and (v) the financial and psychological impact on victims. We provide our data evaluation techniques and findings as follows. 

\subsection{Technical Setup and Filtration}

We conducted both automated and manual checks on 2,570 narratives. We provide detail on automated checks, and filtration as below. 

\BfPara{Heuristics Categorical Filtration} In heuristics-based categorical filtration we created keywords based on random 250 narrative observations in each category. To identify the initial contact method, we applied regex searches for keywords such as \emph{contact}, \emph{app}, or specific social media platform names such as \emph{Twitter}, \emph{Instagram}, \emph{WhatsApp} and others. For relationship-building tactics, we looked for keywords indicating \emph{photo} sharing, \emph{screenshots}, \emph{attachment}, or romance-related words (e.g., \emph{love}, \emph{charm}, \emph{beautiful}, \emph{handsome} as well as investment terms (e.g., \emph{fast}, \emph{return}, \emph{easy}, \emph{crypto}). To detect scammers’ techniques for stealing funds, we searched for indicators of payment methods like \emph{PayPal}, \emph{bank}, \emph{credit card}, \emph{cryptocurrency addresses}, \emph{email}, \emph{URL}, and \emph{phone}. In the categories of emotional and psychological manipulation, we focused on expressions of urgency, false promises, and guarantees. For financial and psychological impacts, we perform searches for keywords such as \emph{loss}, \emph{savings}, \emph{bankruptcy}, \emph{depression}, \emph{anxiety}, \emph{betrayal}, \emph{shame}, and \emph{guilt} which reflect victims’ financial losses and mental health impacts. 

\BfPara{Sub-Categorical and Quantitative Filtration} For three categories—\emph{Relationship-Building}, \emph{Scammer Techniques}, and \emph{Financial and Psychological Impact}—we developed LLM-based prompt queries to identify further sub-categories \cite{openAI}. This included prompts to explore specific relationship-building techniques used by scammers, the scamming tactics applied to victims, and the financial and psychological impacts victims face post-scam. For \emph{Initial Contact} and \emph{Financial Loss Metrics}, we used keyword-based extraction to identify associated values instead of prompt-based querying.

\BfPara{Manual Quality Check} In addition to the two automated checks using heuristics and prompt engineering, we conducted a manual evaluation of the collected data. For \emph{Initial Contact} and \emph{Financial Loss Metrics}, we calculated data values by manually curating each context. For prompt-engineered queries, we manually evaluated 30\%-50\% of the data within each narrative sub-category to ensure quality and relevance. Our analysis demonstrated that LLM-based sub-categorical filtering performed effectively across the three categories, and consistently categorizing relevant contexts with high accuracy.

\subsection{Results} We present data evaluation and metrics for each category within the narrative analysis of the life cycle of pig-butchering scams. This includes illustrating how scams begin, how scammers build relationships, the aftermath for victims, and the overall tactics used by scammers. The details are provided as follows.

\BfPara{Initial Contact} The victim's shared experience mentioned that scammers often initiate contact with the victim through various dating apps, social media platforms, and accidental text messages. We identified 1,593/2,570 victim sharing being reached out to scammers in 16 distinct platforms. These includes: \emph{WhatsApp} (508), \emph{Instagram} (201), \emph{Telegram} (198), \emph{Facebook} (197), \emph{Tinder} (95), \emph{Match} (48), \emph{Hinge} (43), \emph{Signal} (40), \emph{Linked} (36), and \emph{Twitter} (30), \emph{YouTube} (22), \emph{TikTok} (17), \emph{SnapChat} (10), \emph{Discord} (10), \emph{Google Chat} (10), \emph{Plenty of Fish} (18). We identified 110 victims who mentioned getting text without specifying applications or platforms. We suspect the 110 belonging to phone text messages. 

\BfPara {Relationship-Building} We identified 1,427/2,570 victims expressing how scammers build relationships with victims. Based on the analysis, we observe scammers use 10 distinct techniques to build relationships with victims. Among these, the top five techniques are as follows: (i) \emph{Friendship}: scammers often begin as \emph{friends} engaging in casual conversations to build connection (291); (ii) \emph{Romance}: scammers create a romantic atmosphere by discussing dreams of a future together and using affectionate language; (iii) \emph{Trust}: scammers build trust by sharing personal stories, showing empathy, and maintaining consistent communication (278); (iv) \emph{Commitment}: scammers talk about the potential for a committed relationship and promise loyalty (132); and (v) \emph{Future/Connection}: scammers suggest a promising future together, often implying financial security through their connection (74). Examples of victims narratives include: 

\smallskip
\emph{The scammer shared stories about their family and struggles, making me feel they understood me.}

\smallskip
\emph{After talking about the future, the scammer told me that for a better retirement I should invest in Hodlsofltd.com.}

\smallskip

\BfPara{Scammer Techniques} Of 2,570 victims, 1,175 shared scammers' method of operations that often resulted in financial losses. These include (i) \emph{Fake Investment Platforms}: scammer often directing victims to fake websites/apps (320); (ii) \emph{Advance Fee for Withdrawal}: victims are told they must pay fees or taxes to withdraw funds, with new fees added to withdraw each time (275); (iii) \emph{Cryptocurrency Transfer Requests}: scammers instruct victim to buy cryptocurrency and transfer it to specific wallets under the guise to investment (240); (iv) \emph{Fake Customer Support}: scammers impersonate customer support agents who inform victims of account issues, and agent to resolve the issues (190); and (iv) \emph{Romantic Coercion}: scammers use romantic influence, and convincing victims to investment in securing future together (150). Examples of victims narratives include: 

\smallskip
\emph{When I contacted customer support about withdrawals, they told me my account was flagged and needed a deposit to verify my identity}

\smallskip
\emph{He told me that if I invested in crypto with him, we could buy a house together and start a life.}

\smallskip

\BfPara{Emotional and Psychological Manipulation} 1,280 victims shared 5 distinct emotional and psychological manipulation techniques that scammers performed. These include (i) \emph{Love Bombing} - scammer expresses overwhelming romantic feeling with victims (340); (ii) \emph{Guilt-Tripping}: scammers make the victim feel guilty in not trusting the scammer (290); (iii) \emph{Urgency and Pressure}: scammers create false sense of urgency and pressured to make decisions in investments or payments (260); (iv) \emph{Isolation}: scammers discourage victim from discussing the relationship or rewards with others (210); and (v) \emph{Future Promises}: scammers promise a future together, and using dreams of shared goals to deepen the emotional attachment and manipulate victims (180). Examples of victims' narratives include: 

\smallskip
\emph{He said our connection was special and private, convincing me to keep it a secret from my friends and family.}

\smallskip
\emph{We would talk about our future plans, mentioning how investing together would help us buy a house.}

\smallskip

\BfPara{Financial and Psychological Impact} We observe 1,490/2,570 victims shared their life impact related to financial and psychological after being the victim of a scam. In \autoref{table:abuse_db_report_final_loss_amount}, we present a breakdown of financial losses by reported currency type. Among these, the victims reported losses in USD was the highest, totaling approximately \$1.2 million. Overall, losses were reported in five fiat currencies and three cryptocurrencies, with the total loss estimated at over \$5.6 million USD. 

\noindent We conducted an analysis of the financial and psychological impacts experienced by victims following the scam. These include 5 distinct impact types: (i) \emph{Life Quality and Financial Loss}: victims lose money often from their life savings, and retired plans resulting in life quality ruin (500); (ii) \emph{Debt and Bankruptcy}: victims take out loans or go into debt after the scam, and sometimes result into bankruptcy (320); (iii) \emph{Emotional Trauma}: victims experience severe emotional distress, and feeling of betrayal (270); (iv) \emph{Mental Health Issues}: victims report mental health struggle expressing anxiety, fear, depression and suicidal thoughts (220); and (iv) \emph{Social Isolation}: victims feel ashamed and often felt to remain isolated in social connection due to the fear of guilt or judgment (180). Examples of victims narratives include: 

\smallskip
\emph{I borrowed from friends and took out multiple loans. Now, I can’t repay them, and I’m facing bankruptcy.}

\smallskip
\emph{I haven’t told anyone about this because I’m embarrassed and afraid of being judged.}

\begin{table}[tb]
    \small
    \caption{Approximated Dollar amount losses from the disclosed victim from public abuse database reports dataset - This table provides reported losses by victims and includes approximated financial losses, with international currencies and cryptocurrencies. In the last row, \emph{Total} we provide the approximated USD dollar values conversion from the first week of November 2024.}
    \centering
    \begin{tabular}{lrr}
        \toprule
        \rowcolor{gray!0}
        \textbf{Currency} & \textbf{Approx. Loss Value} & \textbf{Victims} \\
        \midrule
        USD	& 1,200,000 & 300 \\
        \rowcolor{gray!10}
        EUR	&  850,000	& 220 \\
        \rowcolor{gray!0}
        GBP	& 600,000	& 180 \\
        \rowcolor{gray!10}
        CAD	& 500,000 & 150 \\
        \rowcolor{gray!0}
        AUD	& 400,000	& 130 \\
        \rowcolor{gray!10}
        USDT & 200,000 & 90 \\
        \rowcolor{gray!0}
        ETH	& 300 & 70 \\
        \rowcolor{gray!10}
        BTC	& 10   & 50 \\
        \rowcolor{gray!0}
        XRP	& 500,000  & 40 \\
        \rowcolor{gray!10}
        \midrule
        Total (Approx. in \$) &  5,631,178 & 1230  \\
        \bottomrule
    \end{tabular}    
    \label{table:abuse_db_report_final_loss_amount}
\end{table}

\section{Analysis of News Outlets}
\label{sec:news_article_evaluation}
In this section, we provide the qualitative analysis of 501 news media coverage articles, identifying 50 case studies of 840 victims of pig-butchering to understand the impact of scams especially to uncover patterns in how often victims are reported, the scale of their financial losses, and the detailed tactics used by scammers. We provide further details of our evaluation as below. 

\BfPara{Study Setup} We conducted two independent studies on 501 news articles. In the first study, 13 junior researchers from our institution were tasked with labeling each article to determine whether it was related to pig-butchering scams. For articles identified as relevant, the researchers further categorized the information into four areas: (i) \emph{News Details}, which included the publication date and type (general, awareness-focused, or victim-related); (ii) \emph{Victim Details}, which captured information such as the victim's name, age, relationship status, financial loss, career or employment status, and the impact experienced; (iii) \emph{Scam Origin and Scammer Details}, covering the platform used, the country of origin, and the fraud type; and (iv) \emph{Authority Involvement}, noting any law enforcement actions, apprehension of fraudsters, or relief provided to victims. After this initial labeling, a second evaluation was conducted by a senior researcher to assess and address any discrepancies in the data. This thorough review required approximately 41.5 hours of analysis in total, averaging about 5 minutes per article for reading and data entry.

\BfPara{Related News and Filtration} In our news analysis, we identified 397 relevant news articles specifically discussing pig-butchering scams. From an initial selection of 501 candidate URLs, 104 articles were excluded after a qualitative review. These included unrelated news (85), paywall-restricted content (5), unavailable content (7), and geographically restricted content (7), which could not be accessed in our study region. Among the 397 relevant articles, 289 were general or awareness-focused, lacking specific details about victim losses or scam operations. Since our study aimed to examine victim losses and the specifics of scam operations, we focused on 108 articles that included identifiable or anonymized victim narratives. Additionally, we performed two filtration techniques on 108 articles to remove duplicate news based on (i) the victim's name, and (ii) the amount of loss reported. Through such filtration, we obtained confirmed 50/108 case studies of news articles on pig-butchering scams. In the following sections, we provide insights based on these 50 news articles' case studies that were extracted from 501 URLs.

\BfPara{Victim Disclosure and Amount Losses} Among the 50 case studies analyzed, 34/50 disclosed victims identities, while 16/50 were reported anonymously. Within the anonymous category, there were two types of cases: (i) 8/16 involved large groups of pig-butchering scam victims, comprising syndicates with 15 to 482 members, totaling 790 anonymous individuals; and (ii) 8/16 represented individual victims or couples who opted to remain anonymous. Of these 50 case studies, 44/50 case studies specified loss amounts, 5/50 did not disclose the loss amount, and 1/50 case study included narrowly avoided being scammed. Using November 2024 exchange rates, the total reported losses across all cases studied amount to approximately \$448,500,944 USD, with an average loss of \$9,750,021, a minimum of \$7,000, and a maximum of \$112,000,000. These figures account for a total of 834 individuals, both with disclosed and anonymous identities while on average the single victim lost \$537,770 in pig-butchering. 

\BfPara{Victim Demographics} Our study on victim demographics includes details on country, sex, occupation, and age group targeted by scammers: (i) \emph{Country}: We identified 33/50 case studies involving victims from the USA, 12/50 provided by anonymous group case studies from various regions (China, Taiwan, Singapore, Australia, and Malaysia), and 3 from India. (ii) \emph{Sex} Among the cases, 11/50 case studies related to male victims, and 6/50 related to female victims specifically. We suspect the anonymous case studies to contain a mix of both male, female or other identified genders. (iii) \emph{Occupation}: We found that 18/50 cases disclosed the victim's occupation, including retired individuals (2), tech/engineering professionals (4), business/real estate professionals (4), and various other fields such as photography, CEO roles, and culinary, which collectively accounted for 8/18 cases. (iv) \emph{Age}: Sixteen cases specified the victim's age range, between 25 to 89 years, with a median age of 51. This shows that pig-butchering scammers strategically target a diverse range of victims across countries, occupations, and age groups, exploiting personal relationships and trust to manipulate victims into significant financial losses.

\BfPara{Psychological and Well-being Impact} We observed that 14/50 case studies mentioned the impact on victims after the scam. Among these, 1 victim tragically committed suicide after losing their life savings and being unable to support their family, 2 cases reported victims filing for bankruptcy, 3 involved victims losing their homes due to bank debts, and 6 victims lost all their life savings, experiencing severe psychological trauma as a result.

\BfPara{Engagement Platform} In 34/50 case studies provided on how scammers initially contacted victims: (i) 17 of these involved social media platforms such as Telegram, Instagram, WhatsApp, and LinkedIn, (ii) 5 were initiated through dating apps (Tinder, Plenty of Fish), (iii) 6 involved social engineering tactics that directed victims to fake crypto trading websites, and (iv) 3 cases reported initial contact through phone or text messaging.

\BfPara{Scam Techniques and Fraud Schemes} We observe 46/50 case studies provided scammers techniques used as part of their fraud schemes. Among these: 20 involved investment or cryptocurrency fraud, 14 were romance-investment scams, and 12 included various other fraud types, such as job fraud, fake mining schemes, wire fraud/SIM-swap, and property scams. This demonstrates that pig-butchering scams extend beyond romance and investment fraud, encompassing a broader range of social engineering tactics.

\BfPara{Authority and Law Enforcement Enagement} We also investigated the role of law enforcement in apprehending these scammers. In 10/50 case studies, details emerged about law enforcement or court involvement, leading to the arrest of 60 scammers connected to cryptocurrency and investment scams, with a total fraud amount of \$172,300,000. These scammers faced charges of conspiracy to commit money laundering, concealing weapons, and fraudulent investment schemes, and authorities uncovered a human trafficking operation from Cambodia, where over 2,000 victims from 11 different countries were being held. Thus, pig-butchering scams are largely sophisticated, syndicate-driven operations, often involving various fraud schemes, money laundering, and even human trafficking networks that exploit victims on an international scale.

\section{Scam Tracking and Financial Loss Metrics}
\label{sec:financial_tracking_and_analysis}
We conducted an automated check to identify fraudsters' associated (i) emails, (ii) URLs, and (iii) crypto addresses within our abuse dataset. This section provides an analysis of these fraudsters' communication channels (emails), external connecting platforms (URLs), and payment methods (cryptocurrency addresses) used in their scam operations.

\subsection{Fraudulent URLs}
From our abuse dataset, we extracted 238 URLs and evaluated them. We provide further details below.

\BfPara{Malicious Check}  Acknowledging that not all were linked to abuse sites, we performed a \emph{VirusTotal} scan to check for signs of maliciousness. This scan identified 66/238 URLs as malicious. To further assess their activity, we used Python's \emph{Requests} library to determine if these URLs were live, revealing that 7/66 were still active. We manually visited these 7 sites, identifying 3 as fraudulent investment websites, 2 as fraudulent crypto-draining sites that prompted users to connect their private keys to fake wallet connectors, 1 domain displaying a \emph{403 Forbidden} message, and 1 as fake tech support site with system infected with virus pop-up notifications and link for a download to scan the system. 

\BfPara{Registered TLDs}
Across these malicious domains, we observed 10 unique top-level domains (TLDs), with the most common being \emph{.com} (48/66), followed by \emph{.cc} (6/66) and \emph{.vip} (2/66). 

\subsection{Fraudulent Email}
From our abuse dataset, we extracted 32 fraudulent emails associated with scammers. 

\BfPara{Registration} Of these, 12 were registered to specific domains, while 20 were created through email providers. These providers included \emph{Gmail} (13), \emph{Yahoo} (2), \emph{Hotmail} (2), \emph{Proton} (1), \emph{AOL} (1), and \emph{iCloud} (1).

\BfPara{Keywords} We analyzed the keywords used in these fraudulent email addresses and identified three patterns: (i) 16/32 contained technical support or crypto-related terms like \emph{support}, \emph{info}, \emph{complaint}, \emph{crypto}, and \emph{service}; (ii) 12 used a username format combining a first and last name; and (iii) 3 included terms associated with government programs, such as \emph{enforcement} and \emph{authority}.

\BfPara{Domain Associated and Active Check} Additionally, we checked 12 unique registered domains associated with these emails, among which four were found to be still active. Upon visiting these domains, we found that one displayed blocked content due to an ad blocker, one redirected to a benign page, one was a classic fake tech support site prompting users to call a listed number, and one was linked to a fraudulent crypto exchange.

\subsection{Fraudulent Cryptoaddresses}
We identified 1,673 crypto addresses in the abuse dataset, of which 1,583 had at least one active transaction recorded on the blockchain. Among these addresses, 3 were Litecoin (LTC) addresses, 749 were Bitcoin (BTC), and the remaining 831 were Ethereum (ETH). We provide our analysis based on transactions from the first week of November 2024, transaction amounts converted to dollar value at the time of each transaction.

\BfPara{Incoming Transactions} These transactions represent the funds received by each account. Collectively, the 1,583 addresses received a total of \$629,339,314, with an average incoming transaction amount of \$397,561. The highest single incoming transaction was \$214,834,563, while the lowest was \$1. Notably, 966 addresses received sums below \$100 in total, whereas two addresses accounted for 68\% (\$429,053,245/\$629,339,314) of the total incoming funds.

\BfPara{Outgoing Transactions} These transactions reflect the amounts sent from each account. The 1,583 addresses sent a combined total of \$380,843,018, averaging \$240,583 per address. A single account contributed 53\% (\$203,103,843/\$380,843,018) of all outgoing transactions. Additionally, 93\% (1,482/1,583) of addresses recorded outgoing transactions totaling less than \$1,000, with an average outgoing amount of \$90.

\BfPara{Creation and Last Active Dates} These addresses were created between 2017 and 2024, with the last activity recorded from 2019 through 2024. Among them, 77\% (1,231/1,583) were created within the past three years, and 39\% (620/1,583) were last active in 2024. We found 193 addresses that transferred out their entire balance after their first transaction, totaling \$6,808,300, with an average outgoing balance of \$35,276.

\BfPara{Disclaimer} Our evaluation of cryptocurrency addresses is based on publicly reported abuse databases from victim reports, and we acknowledge that not all transactions associated with these addresses may be related to scams. However, we consider these addresses to be highly suspicious and likely misused in fraudulent activities targeting victims.

\section{Quantitative Study on Scams}
\label{sec:quantitative_user_study}
We performed a quantitative study of scams where we surveyed a participants from crowdsourcing platform to identify individuals potentially impacted by scams, including \emph{pig-butchering} scams. Our primary goals were to (i) measure the representative online scam victims of pig-butchering scams in comparison to other types of scams and (ii) perform an in-the-wild quantitative assessment of the financial impacts and losses these scams have caused over the past five years. We provide additional detail on survey setup and findings of participants' responses through the hosted survey as below.

\subsection{Survey Setup and Details}

Prior to setting up the survey, we conducted preliminary work to refine various aspects, including survey type, model selection, participant demographics, and ethical considerations. We outline the details of these preparations below.

\BfPara{Representational Study and Target Region} We conducted a representative quantitative study, selecting participants from the United States taking into consideration that pig-butchering scams are higher in the U.S. compared to other regions. Acknowledging the limitations in participant diversity on the crowdsourcing platforms, a prevalence study might not accurately reflect users from multiple countries, so we designed our qualitative study to prioritize representativeness over prevalence.

\BfPara{Questionnaires Model} The survey was structured around the following categories: (i) demographics, covering age group, gender, country of residence, and education level; (ii) financial loss, noting any monetary losses due to scams; (iii) scam type, with further questions on contact methods and social engineering tactics specific to targeted scams; (iv) awareness of scams and knowledge of precautions regarding sensitive information sharing; and (v) additional comments or insights on scams. The complete quantitative questionnaire can be found in Appendix \ref{sec:survey_questionnaires}. 

\BfPara{Survey Hosting and Response Filtration} We created our survey using \emph{Qualtrics}~\cite{qualtrics} and distributed it via the crowdsourcing platform \emph{Prolific}~\cite{prolific} in November 2024. 
A total of 590 responses were received, and we filtered out 6 responses to confirm all participants were from the United States. Our analysis is based on the remaining 584 responses from U.S.-based participants.

\BfPara{Participant Demographics} We ensured all of the participants recruited were from the United States to provide a representative quantitative study. The demographics of participants are as follows: (i) \emph{Age Group}: Participants were distributed across three age groups — \emph{18-24} (278 participants), \emph{25-34} (228 participants), and \emph{35-44} (78 participants); (ii) \emph{Gender}: Participants identified as \emph{Male} (282), \emph{Female} (289), \emph{Prefer not to say} (10), and \emph{Other} (3). (iii) \emph{Education}: The participants’ education levels included \emph{High School} (226), \emph{Bachelor’s} (263), \emph{Master’s} (78), and \emph{Doctorate} (5).

\BfPara{Ethicial Consideration and Data Handling} We consulted our institution’s Empirical Research Group to ensure our survey adhered to ethical guidelines, treated participants with respect, avoided sensitive questions, protected data, and upheld integrity. We did not collect any identifying information, such as names, personal references, or other identifiable data. Throughout the survey, we refrained from sensitive questions, including those potentially causing emotional distress or related to cultural contexts, and all questions were phrased in neutral language. Before beginning the survey, participants received a clear description of the study, procedures, and their expected involvement. They were informed they could withdraw at any time, and contact information for the principal investigator and institutional details was provided for any follow-up inquiries. Each survey participant received a \$1 compensation, and our participants average time spent was 288 seconds (4.8 minutes).

\subsection{Survey Findings}
We analyzed the responses from 584 participants, presenting our findings in this section. Our results focus on eight key insights, detailed below.

\BfPara{Online Scams and Defrauded Victims} In our survey, 46\% (252/584) of respondents reported being victims of online scams or fraud, while 50\% (272/584) indicated they had not been scammed or defrauded in the past five years. Additionally, 10\% (60/584) stated they were unsure if they had been scammed or defrauded.

\BfPara{Scam Categories and Frequency} Of the 272 participants who reported being defrauded, 70\% (191/272) indicated they had fallen victim to a single type of scam across eight categories: (i) \emph{Phishing} (52), (ii) \emph{Fake Online Website} (44), (iii) \emph{Identity Theft} (30), (iv) \emph{Employment or Job Fraud} (17), (v) \emph{Pig-butchering} (7), (vi) \emph{Charity Scam} (4), (vii) \emph{Technical Support Scam} (3), and (viii) \emph{Lottery/Prize Scam} (2). On the other hand, 44\% (120/272) reported being victims of multiple scam types within the past five years. Among these cases, the top three recurring scams were \emph{Phishing} (86/120), \emph{Fake Online Shopping} (54/120), and \emph{Technical Support Scams} (32/120). For participants who experienced repeated scams over the last five years, the reported frequency counts included: two times (67/120), three times (36/120), four times (12/120), and five times (5/120). Additionally, 36 participants described other types of scams, such as credit card theft at gas stations or restaurants, undelivered packages, and phone scams involving fake kidnapping threats.

\BfPara{Amount Lost} In our questionnaires on financial losses from scams over the past five years, users provided responses across various categories: 128 reported no financial loss, 71 lost less than \$100, 77 lost between \$101 and \$1,000, 29 lost between \$1,001 and \$10,000, and 7 reported losses between \$10,001 and \$100,000. Based on these ranges, the total estimated losses fall between approximately \$106,806 and \$1,074,100, with an average loss of \$3,209.

\BfPara{Victims of Pig-Butchering Scams} The participant's responses on whether being a victim of pig-butchering scams within 5 years, indicated that 20 participants had fallen victim to pig-butchering scams. Our specific questions on this type of scam uncovered several details: (i) \emph{Method of Initial Contact}: Victims reported initial contact through various platforms, including dating apps (9), social media (6), and other methods (5), such as emails or cryptocurrency exchange websites. (ii) \emph{Victim Grooming Period}: The time scammers spent building trust varied, with 7 participants reporting a grooming period of 1-2 weeks, 5 reporting 3-4 weeks, 2 indicating 1-3 months, and 5 experiencing over 3 months of interaction. (iii) \emph{Reasons for Financial Loss}: Among the participants, 10 reported losses due to fraudulent cryptocurrency investment websites, 6 due to romance scams, and 4 from other types of fraud, including employment scams, cash giveaways, and bank transfers.

\BfPara{Participants' Awareness of Pig-Butchering Scams} A majority of 78\% (460/584) of participants reported that they had never heard of pig-butchering scams, while only 14\% (85/584) were familiar with this type of scam, and 6\% (38/584) were uncertain. These findings highlight a clear need for increased public awareness and education on pig-butchering scams.

\BfPara{Participants' Awareness of General Scams} Among 584 participants, 312 indicated familiarity with online scams. Of these, 34\% (106/312) reported being moderately familiar, while 35\% (111/312) described themselves as very or extremely familiar. Additionally, 6\% (21/312) were not familiar at all, and 23\% (74/312) indicated slight familiarity with online scams.

\BfPara{Online Precautionary Measures} We asked participants what precautionary steps they take to avoid being scammed, with five main choices as well as an option for open-ended responses. The responses included: regularly monitoring financial accounts for fraud (75), avoiding sharing personal information (79), educating themselves about new scams regularly (61), verifying unknown contacts across multiple platforms (48), avoiding unsolicited investment opportunities (34), following all the listed measures (12), avoiding downloads (1), and not answering calls from strangers (1).

\BfPara{Additional Comments on Online Scams} We asked participants to provide additional comments or thoughts on online scams and received several such comments. We highlight five main such comments: (i) participants recommended increasing public awareness and education on online scams, especially for vulnerable groups like the elderly, (ii) emphasize the importance of verifying information and being cautious with unsolicited messages, (iii) advocate for stronger laws and enforcement to deter scammers, along with proactive security practices like using strong passwords and two-factor authentication, (iv) staying informed on evolving scam tactics, and (v) supporting victims, and encouraging empathy are also suggested as ways to combat the negative impacts of scams on individuals and society. Examples of comments are as below:

\smallskip
\emph{The elder people very vulnerable in these situations. Must be educated by peoples. I always warning my parents about these kinds of scams.}

\smallskip
\emph{Online scams are more common and sophisticated. Stay cautious, verify sources, don’t share personal info, and use two-factor authentication to protect yourself.}

\section{Discussion}
\label{sec:discussion}

In this section, we provide additional detail on the ethical considerations and limitations of our dataset. 

\subsection{Ethical Considerations and Disclosure} Our research adheres to strict ethical standards and consulted the internal Empirical Research Team to ensure that our survey questionnaires, models, and data handling comply with data management and \emph{GDPR} guidelines. Prior to the survey, participants were informed about the research goals and data handling practices. We avoided collecting any personally identifiable information and ensured that data collection was conducted anonymously. Furthermore, data gathered from social media platforms, abuse databases, and news outlets consists solely of publicly reported information, with no direct interaction with victims or scammers. We disclosed the scammer's cryptocurrency addresses involved in scams to \emph{Chainabuse} for further action.

\subsection{Dataset Limitations} Below we provide limitations on our social media, news, and public abuse report dataset.  

\BfPara{Social Media Posts} Our social media data collection was restricted to publicly accessible data that did not require special permissions, memberships, or user-specific relationships to view. We avoided any human interactions during data collection and did not join any social media groups or membership-based communities to collect data. All social media data sources were accessed through APIs.

\BfPara{News Dataset} Our dataset collection relied on news and web searches using three search engines: \emph{Yahoo}, \emph{Bing}, and \emph{Google}. We excluded geographically restricted content and did not crawl data with varied location settings. Consequently, we may have missed news targeted at specific regions or content tailored to particular geographic locations. However, our keyword selection focused on English-language news articles with pig-butchering-specific terms, which means that non-English or region-specific news articles may not be represented in our dataset.

\BfPara{Abuse Dataset} Our public report abuse dataset is based on U.S.-based reports, so the data may not fully represent reports from other languages or regions less familiar with \emph{Chainabuse} and the \emph{Crypto Scam Tracker}. However, we argue that since the U.S. has the highest number of pig-butchering scam victims, these reports likely provide a representative sample of such scams. Although we aimed to collect additional publicly available data for this study, this was not possible. Instead, we collaborated with \emph{Chainabuse}, and the \emph{Crypto Scam Tracker} dataset was publicly accessible, which limited our dataset to these sources. Given that both data sources are leaders in fraud tracking, we believe this data is fairly representative of pig-butchering scams.

\section{Recommendations}
\label{sec:recommendations}
In this section, we provide recommendations for fighting pig-butchering scams based on the insights collected from the findings of our research. We mainly provide to three different entities: (i) social media platforms, (ii) users, and (iii) policymakers. We provide further details below.

\BfPara{Recommendation to Users} We encourage users to be cautious when responding to messages or unsolicited offers. Scammers often initiate contact through direct messages or public post engagements to lure potential victims with targeted schemes such as romance fraud, investment fraud, or similar scams. Sharing private or sensitive information on social media should be done with care, as scammers may exploit this information to build trust and craft convincing stories, ultimately leading users to fall for pig-butchering schemes. Any investment platforms should be thoroughly verified to confirm their legitimacy. Besides, it is important to perform reverse checks on social media users and associated platforms before engaging in any financial transactions.

\BfPara{Recommendation to Social Media Platforms} We recommend social media platforms regularly monitor profile metadata and user engagement solicitations. We urge online platforms, particularly financial services, and social media sites, to take proactive steps in identifying and preventing pig butchering scams and their variances. These actions should include monitoring user profile metadata connected to external fraudulent websites, cryptocurrency addresses with suspicious or flagged transaction histories, suspicious emails, phone numbers, and similar indicators. Such accounts or content should be blocked or flagged with a potential fraud warning to alert users of potential risks. Accounts engaging with unknown or unverified connections should also be closely monitored.  

\BfPara{Recommendation to Policy Makers} As Pig-butchering is a well-planned scam that involves abuse of a multi-layered network, mitigation and safeguarding against such scam requires various policymakers such as government, law enforcement, cybercrime threat intelligence, researchers, social media platforms, and security communities to work collaboratively to tip in any form of suspecting entails. Examples of such regulations include regulating cryptocurrency exchanges, monitoring unusual financial transactions, collaboration with social media platforms, international law enforcement collaboration, streamlining the legal processes such as freezing assets, pursuing criminals, and recovering funds in a timely manner, and ensuring that various sectors are made accountable to fight such scam with regulatory cyber security standards. Through such measures, policymakers can work towards better protecting the potential victims and creating a secure and informed financial environment.

\section{Conclusion}
In this research, we conducted a comprehensive analysis of pig-butchering scams through previously unexplored sources, focusing on social media, abuse report databases, and news outlets. We identified that scammers employ various social engineering techniques to lure victims across different platforms, extending beyond traditional romance and investment fraud. This large-scale study analyzed victim narratives shared across over 430,000 social media accounts, 770,000 posts, 3,200 abuse database entries, and 1,000 news articles. We uncovered a total of 146 social media accounts, 2,570 abuse database narratives, and 50 case studies of 834 victims who collectively lost over \$521 million to pig-butchering scams. Additionally, we tracked fraudulent channels and payment methods scammers directed victims to use. Our quantitative survey on online scams revealed that 50\% of the participants had been defrauded in some form, with 20 sharing specific experiences of pig-butchering scams. Based on these findings, we offer recommendations for platforms, users, and policymakers to create proactive defenses against such scams.

\textbf{}

\noindent \textbf{Acknowledgements.}
We extend our gratitude to Ian Schade (Chainabuse) for providing valuable data and insights related to pig-butchering scams and the associated cryptocurrency accounts. We also greatly appreciate the efforts of the students who assisted with the manual review of news articles: Ady, Sagar, Khalil, Abdullah, Alex, David, Keno, Addison, Pouya, Raphael, and Jonathan. This work was funded by the German Federal Ministry of Education and Research (BMBF grant 16KIS1900 ``UbiTrans''). 

\bibliographystyle{ieeetr}
\bibliography{strings,bib}

\appendix
\subsection{Search Keywords Formation} 
\label{sec:manual_keywords}
In order to identify posts that are relevant to pig butchering, we initially dive deep into a context that is reported as first-hand experience of pig butchering. Based on such observations, we identified for main categories which are explained further below.   

\begin{itemize}

 \item \textbf{Dating/Romance.} We noted that pig-butchering scammers frequently target users seeking dating or romantic relationships through social media. The keywords in this category include phrases such as \emph{find a girlfriend}, \emph{guaranteed sex}, \emph{date Asian}, \emph{mystical romance}, etc. In total, we compiled 49 keywords related to these themes.

 \item \textbf{Investment Fraud.} We observed that social media users frequently fall victim to pig-butchering scams related to investments. These scams involve both traditional investments and cryptocurrencies. Keywords in this category include terms such as \emph{high returns}, \emph{quick gains}, \emph{investment mastery}, \emph{crypto invest}, \emph{double cryptocurrency}, \emph{wealth guard}, and \emph{crypto growth fund}. In total, we identified 100 keywords associated with investment-related fraud cases.

 \item \textbf{Fake Jobs.} We observed that some pig-butchering cases involved scammers offering fake job opportunities. Keywords related to these fake jobs include phrases such as \emph{easy remote job}, \emph{quick job abroad}, \emph{dream careers}, \emph{job security guaranteed}, and \emph{fast track job employment}. In total, we identified 36 keywords associated with fake job scams.

 \item \textbf{Case Studies Track.} We observed that cases related to pig-butchering were often shared as alerts using hashtags. These hashtags/keywords typically included phrases such as \emph{romance scam tracker}, \emph{pig-butchering tracker}, \emph{heartbreak scam warning}, and \emph{fraudulent love alert}. In total, we identified 14 keywords related to popular hashtag case studies.
\end{itemize}

\begin{figure}[tb]
\centering
\includegraphics[width=.40\textwidth]{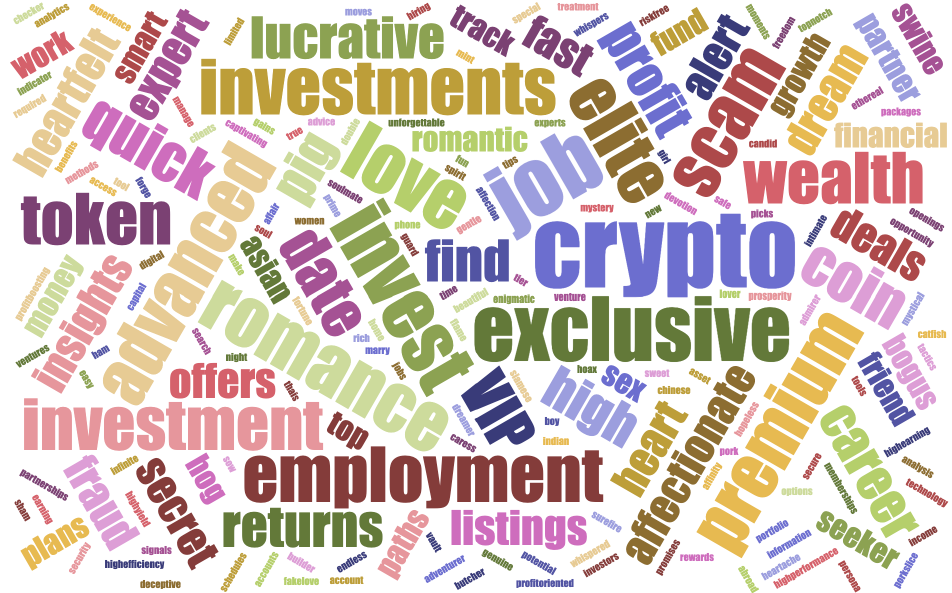}\hfill
\caption{Search keywords word composite: In this figure, we display the word composite used to perform queries for collecting data direct victims of pig-butchering scams. The figure shows that the word composition is higher in contexts related to ~\emph{investments}, ~\emph{crypto}, ~\emph{employment}, ~\emph{quick wealth}, ~\emph{romance}, ~\emph{dating}, ~\emph{love}, and ~\emph{career}. }
\label{fig:word_clouds}
\end{figure}

Thus, based on our manual observation, we created a total of 219 keywords that we used as part of search posts on social media platforms. We provide the word frequency composite in~\autoref{fig:word_clouds}. 

\subsection{Social Media Data Filtration}
\label{sec:social_media_data_filtration}
In this section, we outline the process of crafting prompts to identify posts related to pig-butchering, romance, or investment fraud. We then perform a manual analysis of the responses to verify the effectiveness of this filtration approach. LLMs were selected due to their effectiveness and adaptability in handling a variety of natural language processing tasks, making them particularly suited for accurately classifying fraudulent donation requests. We provide the details of automated and manual filtration below.

\BfPara{Automated Filtration} To determine if a post is related to one of these three cases: pig-butchering, romance scam, or investment fraud, we designed a prompt that evaluates whether the input post includes one of these contexts, outputting the result as a boolean (true or false). Using the OpenAI API~\cite{openAI}, we queried each of the posts of four social media platforms. Below, we provide examples of prompt instruction for \emph{pig-butchering scam} along with input samples for responses received in both cases (false and true). We repeat this process for each post to \emph{romance} and \emph{investment fraud.}

\BfPara{Prompt Instruction}
\begin{mdframed}[style=insightstyle]
\begin{quote}
You are given a text and tasked with determining if it describes a first-hand experience of a pig-butchering scam. The output must be a boolean value, either true or false, formatted as a Python boolean. Provide no explanation.
\end{quote}
\end{mdframed}

\BfPara{Input Sample Post - API Response True Case}
\begin{mdframed}[style=insightstyle]
\begin{quote}
I got scammed by someone claiming to be Drew Barrymore. We had a toured romance through Google chat.. anyway look out much of a bad forgery this is.. @DrewBarrymore
\end{quote}
\end{mdframed}

\BfPara{Output of ChatGPT - API Response True Case}
\begin{mdframed}[style=insightstyle]
\begin{quote}
True
\end{quote}
\end{mdframed}

\BfPara{Input Sample Post - API Response False Case}
\begin{mdframed}[style=insightstyle]
\begin{quote}
Nft and Game are ready to launch as soon as we complete the presale - we will complete the presale in about ... 1 second - so keep your LunarRabbit tokens for gaming experience and great income from activities in the LunarRabbit ecosystem 
\end{quote}
\end{mdframed}

\BfPara{Output of ChatGPT - API Response False Case}
\begin{mdframed}[style=insightstyle]
\begin{quote}
False
\end{quote}
\end{mdframed}

\BfPara{Manual Filtration} In our manual filtration process, we conduct a qualitative analysis of each post flagged as \emph{true} by the LLM prompt response to further assess whether the narratives align with pig-butchering scams. This filtration process specifically identifies instances where scammers groom the victim before committing fraud, as pig-butchering scams often build upon other types of fraud, such as romance and investment schemes, thus distinguishing them from standard romance and investment scams. Additionally, we performed a random evaluation of the subset cases where LLM responses were \emph{false} and identified that classification held correctness with a negative response.  
\subsection{Survey Questionnaires}
\label{sec:survey_questionnaires}
Our survey questionnaire for participants from \emph{Prolific} was structured around six categories. These categories include: (i) Participant Demographics, (ii) General Questions on Scam Experiences, (iii) Specific Scam Encounters, (iv) Focus on Pig-Butchering Scams, (v) General Awareness \& Prevention, and (vi) Closing Questions. The complete list of questions is provided below.

\BfPara{Demographics} In participant's demographics we ask questions related to age group, gender, country of residence, and education level.

\hfill

\noindent\textbf{What is your age group?}
\begin{itemize}
    \item 18-24
    \item 25-34
    \item 35-44
    \item 45-54
    \item 55+
\end{itemize}

\hfill

\noindent\textbf{What is your gender?}
\begin{itemize}
    \item Male
    \item Female
    \item Prefer Not to Say
    \item Other
\end{itemize}

\hfill

\noindent\textbf{What is your country of residence?}\\
\noindent(Text Field)
\hfill

\hfill

\noindent\textbf{What is your education level?}
\begin{itemize}
    \item No schooling
    \item High school
    \item Bachelor's
    \item Master's
    \item Doctorate
\end{itemize}

\BfPara{General Questions About Experiences with Scams}  
We ask participants whether they have encountered online scams within the past five years.

\hfill

\noindent\textbf{Have you been scammed or defrauded in the last 5 years? If "No" or "Unsure," you may want to skip them to the end of the survey (\#6)}
\begin{itemize}
    \item Yes
    \item No
    \item Unsure
\end{itemize}

\hfill

\noindent\textbf{How many times have you been scammed or defrauded in the last 5 years?}
\begin{itemize}
    \item 1
    \item 2
    \item 3
    \item 4
    \item 5
    \item 5+
\end{itemize}

\hfill

\noindent\textbf{How much money did you lose in total due to scams?}
\begin{itemize}
    \item Less than \$100
    \item Between \$101 to \$1,000
    \item Between \$1,001 to \$10,000
    \item Between \$10,001 to \$100,000
    \item Between \$100,001 to \$1,000,000
    \item 1 million+
    \item Did not lose any money
\end{itemize}

\BfPara{Specific Scam Experience}  
We ask participants whether they have encountered one or more types of scams from the provided lists of scams.

\hfill

\noindent\textbf{Which of the following scams have you experienced in the last 5 years? (Select all that apply)}
\begin{itemize}
    \item Phishing (email/SMS)
    \item Fake Online Website/Shopping Scams
    \item Lottery or prize scams
    \item Identity theft
    \item Charity fraud
    \item Employment or job offer
    \item Other (Please specify) [text box]
\end{itemize}

\hfill

\BfPara{Focus on the Pig-Butchering Scam}  
Among the participants who have experienced pig-butchering scams, we ask participants specifics to such scams.

\hfill

\noindent\textbf{If you selected a Pig-butchering scam, please provide a specific experience. How did the scammer initially contact you? (Social media, Dating app, Messaging app, Email, other)}
\begin{itemize}
    \item Social media
    \item FDating app
    \item Messaging app
    \item Email
    \item Other (Please specify) [text box]
\end{itemize}

\hfill

\noindent\textbf{How long did the scammer build trust with you before asking for money or investments?}
\begin{itemize}
    \item 1-2 weeks
    \item 3-4 weeks
    \item 1-3 months
    \item 3 months+
    \item Other (Please specify) [text box]
\end{itemize}

\hfill

\noindent\textbf{What type of investment did they ask you to make? }
\begin{itemize}
    \item Cryptocurrency
    \item Stock/Trading
    \item Real estate
    \item Other (Please specify) [text box]
\end{itemize}

\hfill

\noindent\textbf{Did you report the scam to any authorities? }
\begin{itemize}
    \item Yes
    \item No
\end{itemize}

\hfill

\noindent\textbf{If yes, which authorities did you report it to?}
\begin{itemize}
    \item Local police
    \item Federal Authorities
    \item Bank
    \item Other (Please specify) [text box]
\end{itemize}

\hfill

\BfPara{General Awareness \& Prevention}  
We aks participants on whether they were familiar with online scams, and what kind of precautions do they take as part of preventing such scams.

\hfill

\noindent\textbf{Before you were scammed, how familiar were you with common online scams (e.g., phishing, investment scams)?}
\begin{itemize}
     \item Not Familiar
    \item Familiar
    \item Highly Familiar
\end{itemize}

\hfill

\noindent\textbf{What precautions do you take now to avoid being scammed? (Select all that apply)}
\begin{itemize}
     \item I do not share personal or financial information online
    \item I verify unknown contacts through multiple platforms
    \item I avoid unsolicited investment opportunities
    \item I regularly monitor my financial accounts for fraud I educate myself about new scams regularly
\end{itemize}

\hfill

\BfPara{Closing Questions}  
Finally, we ask participants if they have heard of pig-butchering scams and invite them to share any additional comments or thoughts on online scams.

\hfill

\noindent\textbf{Have you heard of the pig-butchering scam before this survey?}
\begin{itemize}
     \item Yes
    \item No
\end{itemize}

\hfill

\noindent\textbf{Do you have any other comments or thoughts on online scams? }

\hfill

\noindent[text box]

\end{document}